# Personality correlates of key roles in informal advice networks

Battistoni, E., & Fronzetti Colladon, A.

# Personality correlates of key roles in informal advice networks

Elisa Battistoni and Andrea Fronzetti Colladon

**Abstract**

Prior research has emphasised the importance of informal advice networks for knowledge sharing and peer learning. We use Social Network Analysis to detect individuals who play a strategic role in advice networks. Even if roles have been extensively described, how to identify people within them is still an open issue. Furthermore, we investigate whether an association between key players and the big five personality traits exists, by means of nonparametric statistics. To achieve this, we present a case study which involves roughly 180 university students. We found 21 of them playing a key role. Results give evidence of significant associations between key positions and Conscientiousness, Neuroticism and Agreeableness; whereas no evidence was found for a relationship with Extraversion or Openness to Experience. Consistently, personality emerges as a relevant indicator for predicting people who are more likely to play a strategic role, even when connection patterns are unknown.

## 1. Introduction

During the last decades, teaching and learning strategies have changed a lot. More and more, students are learning from each other without direct teacher intervention (Boud, Cohen, & Sampson, 1999). Informal interactions with peers are also predominant in the academic profession (Boud, 1999) and in business contexts (e.g., Boud & Middleton, 2003; Cross, Parker, Prusak, & Borgatti, 2001). Prior research has emphasised the importance of social relationships for acquiring information (e.g., Burt, 1995; Granovetter, 1973) and to learn others' work (Brown & Duguid, 1991; Lave & Wenger, 1991). People's relationships have a significant impact on their capability to learn, to obtain relevant information and to solve problems (Cross et al., 2001). As a result, knowledge and information flows in informal networks emerge as critical resources to be addressed and managed (Grant, 1996). From this point of view, advice-seeking networks are particularly important for they show the prominent players on whom others depend to solve problems (Krackhardt & Hanson, 1993) and they are a means for individuals to share resources such as information and knowledge (Sparrowe, Liden, Wayne, & Kraimer, 2001). Nonetheless, informal networks are not frequently assessed neither in learning environments nor in business organizations – even if managers recognise their critical value (Cross, Prusak, & Parker, 2002). The Social Network Analysis (SNA) is probably one of the most useful approaches in this setting and is increasingly applied in business and learning contexts (Wasserman & Faust, 1994). This methodology analyses social relationships and can describe, explore and understand how complex links work and evolve in time. Each actor is represented as a node in the network and is tied to other actors by one or more specific types of interdependence, such as friendship, information exchange or common interests.

In literature, many authors use SNA to analyse informal networks (e.g., Cross & Parker, 2004; Morton, Brookes, Smart, Backhouse, & Burns, 2004). Some of them, in particular, try to spot key roles taking into account the different kinds of relationship between actors and their positions in the network. Nodes, indeed, do not always have the same level of importance. For example, we could find actors with a critical role for they are in a very central position, linking to many others in and out. Many authors have analysed the topic of measuring centrality (e.g., Bonacich, 1991; Freeman, Borgatti, & White, 1991; Freeman, 1979; Katz, 1953) and also investigated its possible associations with power (e.g., Bonacich, 1987; Gomez et al., 2003; Mizruchi & Potts, 1998). Being in a central position can provide easy access to information or advice and can be helpful for endorsing personal ideas, thanks to the great number of connections and a bigger visibility or influence (Klein, Lim, Saltz, & Mayer, 2004); peripheral nodes are often lacking all these elements. Key players are those actors who exhibit a role which is often critical for the performance of the entire network (Chan & Liebowitz, 2006) and one of the best categorisations is given by Cross and Prusak (2002), who refer to central connectors, peripheral specialists, boundary spanners and information brokers. A main concern is to identify such – or other – key players in networks. Some attempts have been made in this direction in order to define metrics which can be used in static scenarios or considering multiple networks or even their evolution over time (e.g., Brendel & Krawczyk, 2010, 2011; Lu, Li, & Liao, 2012; Pasqualino, Barchiesi, Battistoni, & Murgia, 2012). However, a common problem often arises for many of these metrics: they are developed for specific kinds of relationship or uncommon roles, or require a large set of data to be applied (e.g. pictures of a network at different stages of its evolution). In this paper we start from Cross and Prusak's key roles and propose specific identification approaches – mainly based on existing metrics and mapping techniques – which can overcome some of the above-mentioned limitations: they can be applied to a static picture of the network, do not require multiple sets of data and could be extended to go beyond the single kind of relationship they were conceived and tested on.

Nevertheless, in real contexts we face many situations where we are not provided with sufficient information to map an informal network and therefore we cannot use any metric to spot key players; in this case, we have to look for other attributes characterising people who are involved in informal interactions.

Personality was revealed to be an important driver for academic performance (e.g., Busato, Prins, Elshout, & Hamaker, 2000; Chamorro-Premuzic & Furnham, 2003), creativity (Batey, Furnham, & Safiullina, 2010; Furnham & Bachtiar, 2008; Shalley & Gilson, 2004) and innovation behaviour (Amo & Kolvereid, 2005); it was also shown to be linked to some of the key roles (e.g., Burt, Jannotta, & Mahoney, 1998; Klein et al., 2004; Williams, 2002) even if the association of enduring personal characteristics with network positions has only rarely been examined. According to Williams (2002, p. 110) – for example – boundary spanners are 'characterised by their ability to engage with others and deploy effective relational and interpersonal competencies' and they are also creative, innovative, reliable and tolerant people. In our paper, we try to find a relationship between the role played by a person in a network and some combinations of his/her personality traits. In assessing personality traits, we refer to the five-factor model of personality, which is a widely used construct, has gained acceptance as a general taxonomy (Judge, Bono, Ilies, & Gerhardt, 2002) and trascends language and other cultural differences (Yamagata et al., 2006).

## 2. Key roles in informal advice networks

When looking for key players in a social network it is essential to know how the social interdependencies between actors have been mapped. Since we analysed an information network in a learning environment, an incoming arrow indicates that someone is being asked for information or advice and an outgoing arrow that someone is asking for information or advice (Pasqualino et al., 2012). When analysing study relationships some people might be much more influential in giving advice or sharing knowledge than others; therefore arcs have to be valued. We consider a directed graph where each node represents an actor in the advice-seeking network; so there is an arc from actor $a$ to actor $b$ if $a$ is asking $b$ for advice. Every arc is then assigned with a value depending on the frequency of the interaction: the more advice or knowledge $a$ receives from $b$, the higher the weight of the arc from $a$ to $b$. Not every arc is bi-directed and the act of giving advice or sharing knowledge is not necessarily symmetrical. In such a network, we want to look for the four types of key players described by Cross & Prusak (2002). These roles were identified by analysing more than fifty large organizations (Cross & Prusak, 2002) and many studies refer to them when dealing with advice, knowledge or information networks (e.g., Chan & Liebowitz, 2006; Pasqualino et al., 2012; Whelan, Collings, & Donnellan, 2010). The identification of key students in learning environments can be vital for the optimisation of information flows and for the improvement of the general performance.

### 2.1 The Central Connector (CC)

Central connectors are the people with the highest number of social links in networks, those who know who can provide critical information and those who almost everyone in the group talks to. Even if their role is not always recognised by the formal organisational chart, they are highly valuable knowledge resources upon whom network performance relies (Chan & Liebowitz, 2006). They are also often valuable in indirectly connecting other actors (Cross & Parker, 2004; Otto & Simon, 2008). In any case, some of them may end up creating bottlenecks and hold back the informal network. Moreover, being a CC is usually very time consuming. In a learning environment CCs are valuable students, with many direct contacts, who can often help their peers find the information or advice they are looking for.

When searching for CCs, we focus on actors with high values of in-degree and out-degree – so linking to many other nodes in and out. In addition, total ties weight must be significant when compared with other nodes (the weighted degree of a vertex is defined as the sum of the weights of its incident edges). Each arc weight represents the frequency of interaction between two nodes (in one direction), when giving or receiving advice. To sum up, we search for actors who are reached by many arcs – balanced inward and outward – and who frequently interact with others, exchanging a significant amount of knowledge. Consistently with this view, we refer to degree centrality measures (Freeman, 1979) in order to identify CCs. Threshold values for weighted in- and out-degree have to be set by the analyst, so as to identify the most central people. This values have to be chosen considering the specific kind of relationship mapped and the network density. In fact, when a graph is highly connected with many strong ties there may be no need to isolate the most central people; even removing some of them, indeed, the information flows would probably not be compromised. In our case study, an appropriate threshold is the 75$^{th}$ percentile for both in- and out-degree measures.
We also excluded those actors who are not connected to at least three colleagues.

**2.2 The Boundary Spanner (BS)**
Boundary spanners are people who nurture connections outside their informal network, serving as bridges with other communities, departments or even organisations. They go beyond their personal affiliations and play a vital role in exchanging experiences and knowledge and in building up strategic alliances. They are in a very good position for recognising innovation opportunities (Chan & Liebowitz, 2006), favouring intergroup transactions and managing intergroup conflicts (Callister & Wall, 2001). In learning environments it may be easy to find subgroups in which the interaction is more intense and BSs are those students who can facilitate communication flows between different cohesive clusters (Pasqualino et al., 2012) – so as to create the conditions for trust and interdependencies.

Before identifying BSs we need to isolate the communities they belong to. The problem of finding communities is considered a data clustering problem: it can be solved by assigning each node to a cluster, in a meaningful way (De Meo, Ferrara, Fiumara, & Provetti, 2011). In our data, we were not given an already clustered network – as in the case, for example, of business departments – so we used the VOS mapping technique (Eck, Waltman, & Berg, 2010) to find communities; this choice is not mandatory and we recommend using the method that best fits the specific network analysed, also taking into account that the number of BSs is sensitive to the number of communities identified. As a second step, we identified BSs as those actors who connected their cluster to others.

**2.3 The Information Broker (IB)**
Information Brokers are people who have a huge importance in making information flow in the network as they keep different subgroups linked together. Even when they have only a few direct connections, they can have the same power as CCs, which comes from the preservation of connectivity. In fact, removing an IB would split the network into two or more smaller and less effective segments (Chan & Liebowitz, 2006), facing the possibility of interrupting important information flow. At the same time, IBs can share the same power as BSs in shaping diffusion patterns among different subgroups.

IBs can be easily identified by searching for cut-vertices and focusing on those keeping together subgroups of at least three nodes – so we excluded the cases where we would have lost only isolated nodes, had we removed a cut-vertex.

## 2.4 The Peripheral Specialist (PS)

In a relationship network there are people who, for different reasons, are not very connected with others and lie on the periphery: they could fit into the network badly, or they could be new entries. It is also possible that someone chooses to be in that position: this is often the case with experts, who have valuable information and are sought by others if needed. They can be valuable resources as they often possess unique skills and perspectives and can provide their colleagues with specific knowledge (Chan & Liebowitz, 2006; Cross & Parker, 2004). PSs may be experienced students who are usually very well prepared on the topics of a specific course; for this reason, they are frequently contacted by their peers who want to access their specific knowledge. By contrast, they seldom look for the help of their colleagues.

In order to locate PSs we, again, consider the degree centrality measures (Freeman, 1979) and focus on those nodes with high values of weighted in-degree and low values of weighted out-degree – so on people who are frequently sought by others, even if they rarely ask for advice. In our case study, we chose the 75$^{th}$ percentile as a threshold value to define high weighted in-degree and the 35$^{th}$ percentile as the maximum allowed value for low weighted out-degree. We also excluded those actors who are not contacted by at least two colleagues.

## 3. Personality traits and network positions

In some cases, we cannot easily identify key players because we are unaware of the connection patterns among actors. For this reason, we look at the personality traits of the actors, since we believe they can be considered important signals which may help identify key players.

There has been little work describing how individual differences affect social structures, and only a few studies investigating the association between personality and network positions (e.g., Burt, 2012; Gloor et al., 2011; Kalish & Robins, 2006). Research by Casciaro et al. (1999) shows how people's perception of their local network is influenced by their personality. The structural approach emphasises the importance of being at the right place (Brass, 1984), but ignores the possibility that the network position might be influenced by the actors' psychology. Other studies, by contrast, give more insight into the origins of network positions and the importance of individual characteristics. Among them, Burt et al. (1998) claim that personality varies with structural holes and find associations between network constraint and personality; Kadushin (2002) sustains that an individual's object-relations orientation in early childhood leads to network cohesion and low network constraint in adulthood; Hallinan and Kubitschek (1988) analyse students' friendship networks and explain how those with a high in-degree centrality are more intolerant to intransitive triads and tend to remove intransitivity over time; Kalish and Robins (2006) find that psychological predispositions are important factors in the formation of ego-networks; Mehra et al. (2001)

state that personality affects social structure, as they find that the longer high self-monitor people stay in an organization, the more they tend to occupy positions of high betweenness centrality. Moreover, they show how personality, together with centrality, is predictive of the workplace performance of individuals. Klein et al. (2004), in addition, state that individuals' enduring personal characteristics influence their acquisition of central positions – as an example they find that education and emotional stability are key predictors of centrality. They also find that people who are low in Neuroticism tend to have high degree centrality scores in advice and friendship networks.

Consistently with these views, we believe that psychological predispositions may add explanatory capacity to network analysis; so we consider personality in order to locate key players and we refer to the five-factor model that suggests five traits which can be used to describe the key personality aspects. These traits are Extraversion, Agreeableness, Conscientiousness, Neuroticism and Openness to Experience. Extraversion is the extent to which a person is willing to engage with the external world. Extraverts are outgoing, gregarious and action-oriented. Agreeableness describes the tendency to be cooperative, friendly, generous and helpful. Conscientiousness refers to the aptitude of being persistent, responsible, and organised. Neurotics are emotionally reactive people and are likely to experience anxiety, anger and depression. Finally, Openness to Experience is the tendency to be imaginative, creative and open-minded. These traits are unaffected by external influences (Asendorpf & Wilpers, 1998) and remain stable throughout a person's lifetime (Allport, 1962; McCrae & Costa, 2002).

According to this introduction, we expect some traits to influence behaviour, and therefore the roles, assumed by actors in the network. Nevertheless, the contribution of each trait will be different depending on the specific kind of relationship mapped. Conscientious people are perceived by others as intelligent, so we expect Conscientiousness to play a major role in advice-networks being positively related with the key roles with the most connections and interactions (CCs and BSs). This is because advice-seeking has a cost and individuals are most likely to connect with people they consider disciplined, experts and hard workers (Borgatti & Cross, 2003; Hinds, Carley, Krackhardt, & Wholey, 2000). Also peripheral individuals who possess unique skills or expertise (PSs) are expected to report high Conscientiousness scores. Conversely, neurotic people tend to be negatively viewed by others (Henderson, 1977) and tend to be less socially supported (Furukawa, Sarason, & Sarason, 1998). Consequently, we hypothesise Neuroticism to be negatively associated with the most connected and frequently contacted key players, and with those keeping together separate portions of the network (IBs). With regard to Extraversion and Agreeableness, we assume actors with high scores in these traits to be friendly and other-oriented, so we also expect these traits to be positively related to those roles that show the highest number of connections and the most heterogeneous ties (CCs, BSs and possibly IBs). For similar reasons, we expect Extraversion to be negatively related to peripheral specialists. Table 1 summarises our hypotheses.

**Table 1**

Expected relationships between key roles and personality traits.

| |
|---|
| Central Connector |
| H1. Conscientiousness is positively related to the role of Central Connector in advice networks. |
| H2. Neuroticism is negatively related to the role of Central Connector in advice networks. |
| H3. Agreeableness is positively related to the role of Central Connector in advice networks. |
| H4. Extraversion is positively related to the role of Central Connector in advice networks. |
| Boundary Spanner |
| H5. Conscientiousness is positively related to the role of Boundary Spanner in advice networks. |
| H6. Neuroticism is negatively related to the role of Boundary Spanner in advice networks. |
| H7. Agreeableness is positively related to the role of Boundary Spanner in advice networks. |
| H8. Extraversion is positively related to the role of Boundary Spanner in advice networks. |
| Information Broker |
| H9. Neuroticism is negatively related to the role of Information Broker in advice networks. |
| H10. Agreeableness is positively related to the role of Information Broker in advice networks. |
| H11. Extraversion is positively related to the role of Information Broker in advice networks. |
| Peripheral Specialist |
| H12. Conscientiousness is positively related to the role of Peripheral Specialist in advice networks. |
| H13. Extraversion is negatively related to the role of Peripheral Specialist in advice networks. |

## 4. Case study

We present a case study applied to the advice-seeking network of our university course (Engineering Economics).

### 4.1 Data Collection

Our class was made up of about 180 students who were asked to complete an online survey to map their interactions in looking for advice and explanations on the course topics. As regards to this kind of relationship, every student had to generate a list of contacts, also reporting the frequency of interaction, on a ten-point scale (with the highest value corresponding to the highest level of interaction). At the same time, in order to assess the personality traits of each respondent, we used the International Personality Item Pool[1] (Goldberg et al., 2006). The response rate was of about 74%; among respondents we excluded those who were not attending our lectures (about 28%) – but just sitting the final exam – because they were not interacting in the classroom. About 83% of respondents were male and 17% females. Age ranged between 19 and 27, with a mean of 20.51 years. Students claimed to be connected, on average, to 1.29 peers (ranging from 0 to a maximum of 6 connections). Frequency of interaction on existing links was rather high – 7.12 over 10, on average, with a standard deviation of 2.51– revealing that students frequently collaborated with colleagues who were among their direct connections. Descriptive statistics for personality traits are presented in Table 2.

---

[1] Specifically, we used the 120-Item version of the IPIP-NEO available at: http://www.personal.psu.edu/faculty/j/5/j5j/IPIP/ipipneo120.htm. This test has demonstrated to be a reliable and valid measure of the five personality factors (Johnson, 2011). For each personality trait, it both provides a numerical score – as percentile estimate – and an ordinal classification of the score (low/average/high), according to whether the respondent is approximately in the lowest 30%, middle 40%, or highest 30% of scores obtained by people of the same sex and age class.

**Table 2**

Descriptive statistics for personality traits.

| Personality Trait | M | SD | Max | Min |
|---|---|---|---|---|
| Extraversion | 56.44 | 22.31 | 94 | 0 |
| Agreeableness | 55.62 | 23.12 | 94 | 0 |
| Conscientiousness | 58.72 | 26.41 | 97 | 0 |
| Neuroticism | 34.93 | 22.33 | 85 | 0 |
| Openness to Experience | 40.16 | 26.49 | 99 | 0 |

We constructed the interaction patterns consistently with the modus operandi described at the beginning of Section 2. Our network emerged to be rather sparse with a density of 0.014, an average degree of 2.59 and about 22% of isolated nodes. Moreover, many students were enclosed in little isolated clusters, except for one big partition standing out with 45 nodes.

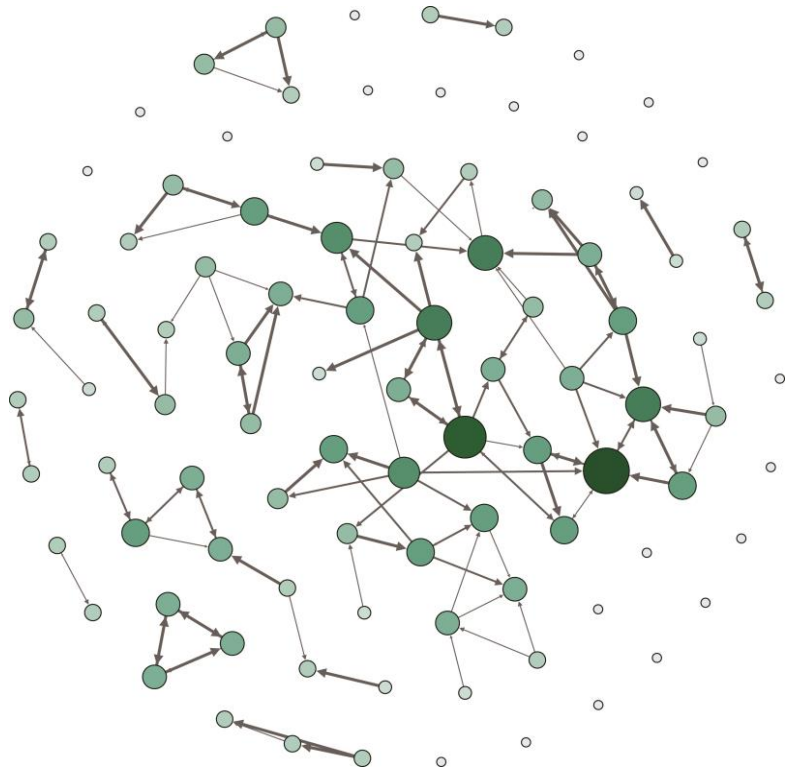

**Figure 1**. Informal advice network of the course of Engineering Economics.

## 4.2 Results

Thanks to the data collected we can use the approaches described in Section 2 in order to identify those actors who are in a key position. In Table 3, we give descriptive statistics for SNA indicators.

**Table 3**

Descriptive statistics for network metrics.

| Network metric | M | SD | Max | Min |
|---|---|---|---|---|
| In-degree | 1.29 | 1.30 | 6 | 0 |
| Out-degree | 1.29 | 1.29 | 6 | 0 |
| All-degree | 2.59 | 2.15 | 10 | 0 |
| Weighted in-degree | 9.29 | 9.87 | 48 | 0 |
| Weighted out-degree | 9.29 | 10.20 | 50 | 0 |
| Weighted all-degree | 18.59 | 16.46 | 72 | 0 |

Analysing SNA metrics, we find 21 key players with 10 of them playing a multiple role. For example, node 33 is both CC and BS (linking to 3 other communities). The actors who assume the highest number of roles are node 48 – who is at the same time CC, IB and BS – and node 85 – who is IB, BS and PS. BS and CC are the roles that most frequently coexist; CC and PS, instead, are the only roles which, by definition, cannot coexist. It is worth noting that playing multiple roles can increase the strategic value of actors.

In order to investigate the possible associations between roles and personality traits we use nonparametric statistics. In fact, our sample is relatively small and not fully appropriate to support parametric tests' assumptions; consequently, a better approach is to use methods that do not make assumptions about how values in the population are distributed. First of all, we perform the Spearman's rank correlation between personality traits and role indexes (these indexes, described in Section 2, are dichotomous: they indicate if a social actor plays a specific role or not). In Table 4, we indicate the results.

**Table 4**

Spearman's rank correlation.

|  | Key Role | | | |
| --- | --- | --- | --- | --- |
| Personality Trait | IB | CC | PS | BS |
| Extraversion | .14 | .07 | .02 | -.04 |
| Agreeableness | .17 | .16 | .16 | .46** |
| Conscientiousness | .18 | .31** | .29** | .52** |
| Neuroticism | -.29** | -.35** | -.12 | -.47** |
| Openness to Experience | -.01 | .06 | -.07 | .10 |

** $p < .01$.

As the Table shows, some correlations are significant: Agreeableness is positively correlated with the role of BS, Conscientiousness with CC, BS and PS, Neuroticism negatively with CC, IB and BS; conversely, not a single role is correlated with Extraversion and Openness to Experience. In such a scenario, Conscientiousness and Neuroticism stand out as the most important traits, being correlated to three out of four roles.

To extend our first results, we also perform $\chi^2$ and Mann-Whitney's U tests. $\chi^2$ provides information about the association between two variables without, however, giving its direction; this limit can be overcome thanks to Goodman and Kruskal's cograduation index ($\gamma$), that we also calculate. In cases where expected numbers in $\chi^2$ tables are lower than 5, we use Fisher's exact test in lieu of $\chi^2$ test (Freeman & Halton, 1951). Finally, Mann-Whitney's U is used to check whether the distributions of personality traits significantly differ for key roles. In Table 5, we illustrate the results of our further investigation.

**Table 5**

Nonparametric tests.

| Personality Trait | Statistic | Key Role | | | |
|---|---|---|---|---|---|
| | | CC | IB | BS | PS |
| Extraversion | U | .68 | 1.38 | -.39 | .22 |
| | FI | .61 | 1.62 | .74 | .41 |
| | γ | .16 | .67 | -.14 | .30 |
| Agreeableness | U | 1.59 | 1.60 | 4.46** | 1.53 |
| | FI | 2.36 | 2.17 | 18.48** | 2.17 |
| | γ | .44 | .73 | .87** | .73 |
| Conscientiousness | U | 3.00** | 1.77 | 5.04** | 2.84** |
| | FI | 11.48** | 1.63 | 31.17** | 4.67* |
| | γ | 1.00** | .68 | 1.00** | 1.00* |
| Neuroticism | U | -3.36** | -2.82** | -4.57** | -1.13 |
| | FI | 10.08** | 4.42 | 19.88** | 1.37 |
| | γ | -1.00** | -1.00* | -.94** | -.60 |
| Openness to Experience | U | .57 | -.06 | .93 | -.71 |
| | FI | 1.53 | .59 | 2.46 | .93 |
| | γ | .21 | -.29 | .15 | -.03 |

*Note.* FI = Fisher's exact test statistic; U = Mann-Whitney U test statistic; γ = Goodman-Kruskal gamma coefficient.

\* p < .05.

\*\* p < .01.

It is worth noting that, since ours is an advice-seeking network in a learning environment, Conscientiousness and Neuroticism, as expected, have major impacts in driving key roles. We also checked for a possible association between respondents' ages and key positions without any significant result – this being also due to the fact that respondents' ages are mostly centred around the mean value. Results provide full support for hypotheses 1, 2, 5-7 and 12; H9 is fully supported with the only exception of Fisher's exact test reporting a p-value of .11; by contrast, we find no significant associations to support hypotheses 3, 4, 8, 10, 11 and 13. This is probably a consequence of the specific kind of relationship mapped; Extraversion, for instance, could be more relevant when analysing friendship networks. Openness to Experience does not significantly relate to any of the key roles; Agreeableness seems to be of some importance only for people who go beyond their personal affiliations and nurture connections with two or more different communities (BSs); a high Conscientiousness and a low Neuroticism are sufficient for individuals to assume a central position in their own cluster (CCs).

Results can be of great importance when we are unaware of a network structure, but we have the personality profiles of the actors involved. In particular:

1. We can look for CCs focusing on those people who are low in Neuroticism and high in Conscientiousness; when actors reveal a high Conscientiousness and a low Neuroticism to others, they tend to be perceived as intelligent and are contacted frequently;
2. IBs are likely to be low in Neuroticism;
3. BSs are positively related to high scores in Agreeableness; this makes actors more willing to interact with others and more likely to link with other clusters. In addition, the BS profile emerges as similar to that for CC (presenting a stronger association with high Conscientiousness and low Neuroticism) and, in our graph, about 41% of BSs are also CCs. This is probably due to the fact that people who are highly connected are more likely to have bridging ties;
4. PSs are only high in Conscientiousness, whereas no association emerges with other traits: this is consistent with their peripheral position and their task of providing others with specific knowledge.

As a further step, we checked for an association between personality traits and the inclusion in the biggest network partition (Table 6). This association is significant for Conscientiousness, Agreeableness and, partially, for Neuroticism, indicating that, in an advice-seeking network, conscientious people, with high Agreeableness and low Neuroticism, are more likely to group into bigger and more connected clusters.

**Table 6**

Inclusion in the biggest partition and personality traits.

| Personality Trait | $r_s$ | $\chi^2$ (df=2) | FI | $\gamma$ | U |
|---|---|---|---|---|---|
| Extraversion | -.14 | 1.63 | | -.20 | -1.34 |
| Agreeableness | .21* | 7.42* | | .30 | 2.04* |
| Conscientiousness | .42** | 14.88** | | .61** | 4.05** |
| Neuroticism | -.24* | | 3.59 | -.33 | -2.36* |
| Openness to Experience | -.13 | 3.98 | | -.20 | -1.23 |

*Note.* $r_s$ = Spearman's rho coefficient; $\chi^2$ = Chi-square test statistic; FI = Fisher's exact test statistic; U = Mann-Whitney U test statistic; $\gamma$ = Goodman-Kruskal gamma coefficient.

\* $p < .05$.

\*\* $p < .01$.

Finally, we tested the explanatory value in personality with regard to students' performance: we found a significant correlation of high grades in the final exam of the course (grades are expressed on a continuous scale, ranging from 0 to a maximum score of 30) with high Agreeableness ($r_s$ = .34, p < .01), high Conscientiousness ($r_s$ = .45, p < .01) and low Neuroticism ($r_s$ = –.50, p < .01). These results are also confirmed by Fisher's exact test and by the index γ. Since we previously show an association of key roles with personality traits, it is also interesting to extend this investigation to the possible associations between actors' performance and key roles – key roles, indeed, are linked to personality profiles, more than to single traits. Our further analysis is presented in Table 7.

**Table 7**

Performance and network positions.

| Key Role | Performance | | | $\chi^2$ (df=2) | FI | γ | U |
| --- | --- | --- | --- | --- | --- | --- | --- |
| | M | SD | $r_s$ | | | | |
| CC | 27.88 | 4.22 | .51** | | 11.89** | .93** | 3.89** |
| IB | 15.75 | 6.72 | .03 | | 2.32 | .03 | .21 |
| BS | 24.63 | 5.63 | .59** | | 18.91** | .89** | 4.52** |
| PS | 16.50 | 3.91 | .07 | | 3.64 | .03 | .52 |
| Included in the biggest partition | 18.67 | 9.38 | .42** | 12.11** | | .68** | 3.20** |

*Note.* $r_s$ = Spearman's rho coefficient; $\chi^2$ = Chi-square test statistic; FI = Fisher's exact test statistic; U = Mann-Whitney U test statistic; γ = Goodman-Kruskal gamma coefficient.

** p < .01.

Results show that being very well connected in the advice network repays in terms of performance: CC and BS are the roles which report the strongest associations with a high final score. This is probably due to the fact that being a central connector entails exchanging a significant amount of advice and knowledge, so as to gain a deeper understanding of the course topics; being a boundary spanner, moreover, gives actors the possibility to access several types of knowledge – due to the interaction with people and groups with different expertise. Prior research has emphasised the benefits of bringing many areas of specialized knowledge together (e.g., Demsetz, 1988). On the other hand, being a peripheral specialist, or an information broker, seems insufficient to achieve a good final score. Thus, PSs may be proficient in their specific topic, but still be missing a broader knowledge which comes from interaction with many other colleagues – so from a more central position. As a consequence, being included in a bigger and more connected cluster also shows to be positively associated with performance, even if not as much as being in a central or boundary position.

Nonetheless, our conclusions on key roles predicting performance should be additionally investigated to see if academic outcomes are mainly influenced by the peculiarities of the social role played or by personality traits. What we present here are just preliminary results intended for further extensions.

## 5. Discussion and conclusions

'Individuals forge a texture of relationships that underpins and shapes their future work and outcomes' (Cattani & Ferriani, 2008, p.825), so a focus on their informal relationships is an important addition to that part of social research which is primarily concerned with attributes of social units (Wasserman & Faust, 1994). In learning environments, a large part of the real information flows go through informal networks – students do not learn only from books or lectures. So, taking care of spontaneous relationships becomes vital to monitor and favour informal ways of learning and to improve general education.

Advice-seeking networks are valuable because of the high amount of information and knowledge that is typically exchanged through them; moreover, they can easily be found in many learning environments or business contexts. The case study we carried out considers the advice-seeking network of a learning environment; nonetheless our methodology can be used regardless of the specific setting.

Our main concern is to investigate the characteristics of key players. CCs enrich the network with their presence and usually favour the spreading of information (Pasqualino et al., 2012). BSs are valuable as they serve as bridges between different groups – also across the boundaries of their primary setting, when acting as gatekeepers (Fleming & Marx, 2006); BSs could, for example, be sharing knowledge with colleagues from other courses or universities. The development of this role can be improved by campus activities, exchange programs, or, in business contexts, by management moving employees across business departments; it will be worth choosing those who are more inclined to building social ties – like CCs with a high level of Agreeableness. IBs are also important, but their number should be limited because it is risky to make the network connectivity dependent on single actors; therefore the development of more than one bridging tie between clusters should be encouraged. Lastly, for peripheral people it is important to distinguish between those with integration problems and those who are peripheral by choice – and perhaps also 'specialists' if possessing specific and valuable knowledge. Nonetheless, preserving and improving the connectivity of an advice-seeking network is also wise in order to support and ease information flow and improve the general performance. To this aim we suggest choosing employees with high Conscientiousness, low Neuroticism and high Agreeableness.

In Section 2, we described key players and suggested how to identify them; however, in real contexts, we are often not provided with the informal network structure, so we are not aware of the connection patterns among actors. In this case, the associations we found between the key roles and personality traits are useful. Thanks to our results, we can support the hypothesis of personality traits being potential antecedents of local network structure and we are able to detect those actors who need a deeper investigation for they could be in a strategic position: CCs present high scores in Conscientiousness and low scores in Neuroticism; IBs associate with low scores in Neuroticism; BSs show high scores in Agreeableness and Conscientiousness and low scores in Neuroticism; PSs score high in Conscientiousness. Our findings partially support Klein et al. (2004) in having Neuroticism negatively associated with centrality in advice networks. However, we also found Conscientiousness to be significantly high for central connectors, thus giving evidence of an association that was not previously confirmed by Klein et al. (2004). Consistently with Mehra et al. (2001) we found that the most strongly connected actors (CCs) and those who present bridging ties among different communities (BSs) are positively associated with performance.

Finally, we claim that the emergence of strategic roles can be promoted by placing the right people in the right places. At the same time, we believe our research should be extended further, in order to consider: a larger sample of people with different ages and social backgrounds; new kinds of interdependencies, such as friendship; other social roles (e.g., Gould & Fernandez, 1989). Further research could test the hypothesis that being aware of actors' personality is helpful in predicting how the informal advice network will evolve over time and what the general performance will be. Moreover, it would be interesting to investigate if an association exists between key roles and other elements, such as creativity, intelligence or more specific psychological constructs.